\documentclass[pre,superscriptaddress,twocolumn]{revtex4-2}
\usepackage{amsfonts,amsmath,amssymb,mathtools,graphicx, float,blkarray, bigstrut}
\usepackage[colorlinks=true,allcolors=blue]{hyperref}
\usepackage[svgnames,dvipsnames,x11names]{xcolor}
\usepackage{tikz}
\usepackage{tikz-cd}
\usepackage{multirow}
\usepackage{placeins} 

\graphicspath{{Figures/}}

\def\ie{\rm{i.e.}}
\def\eg{\rm{e.g.}}

\begin{document}

\title{Breadth, Depth, and Flux of Course-Prerequisite Networks}

\author{Konstantin Zuev}
\email[Corresponding author, email: ]{kostia@caltech.edu}
\affiliation{Department of Computing and Mathematical Sciences, California Institute of Technology, Pasadena, CA 91125, USA}

\author{Pavlos Stavrinides}
\affiliation{School of Computational Science and Engineering, Georgia Institute of Technology, Atlanta, GA 30332, USA}

\begin{abstract}
Course-prerequisite networks (CPNs) are directed acyclic graphs that model complex academic curricula by representing courses as nodes and dependencies between them as directed links. These networks are indispensable tools for visualizing, studying, and understanding curricula. For example, CPNs can be used to detect important courses, improve advising, guide curriculum design, analyze graduation time distributions, and quantify the strength of knowledge flow between different university departments. However, most CPN analyses to date have focused only on micro- and meso-scale properties. To fill this gap, we define and study three new global CPN measures: breadth, depth, and flux. All three measures are invariant under transitive reduction and are based on the concept of topological stratification, which generalizes topological ordering in directed acyclic graphs. These measures can be used for macro-scale comparison of different CPNs. We illustrate the new measures numerically by applying them to three real and synthetic CPNs from three universities: the Cyprus University of Technology, the California Institute of Technology, and Johns Hopkins University. The CPN data analyzed in this paper are publicly available in a GitHub repository.
\end{abstract}


\maketitle

\section{Introduction}
\label{sec:Introduction}

Complex systems, consisting of many interconnected and interacting components, are embedded in the material and informational fabric of the modern world. One approach to understanding complex systems is based on modeling a system by a graph (or network), which is a collection of nodes connected by links, that captures the pattern of connections between the system's components and represents its structural skeleton. This natural idea coupled with advances in computing power and statistical analysis of real network data has led to a new interdisciplinary field, \textit{network science}~\cite{Newman2018,NBW2006,Dorogovtsev2010,Easley2010}, which emerged at the intersection of graph theory, computational statistics, computer science, and statistical physics. Networks were used to model and study different technological, information, biological, and social systems such as the Internet~\cite{Faloutsos1999,Pastor-Satorras2004}, power grids~\cite{Arianos2009,Pagani2013}, the world wide web~\cite{Broder2000}, citation networks~\cite{Radicchi2012}, food webs~\cite{Martinez1991}, protein interactions~\cite{Jeong2001}, social groups of people~\cite{Zachary1977,Borgatti2009} and animals~\cite{Lusseau2003}, and even the universe~\cite{Krioukov2012,Cunningham2017} and brain~\cite{Krioukov2014}. In this paper, we focus on the study of course-prerequisite networks~\cite{Stavrinides2023}, a class of information networks that model complex academic curricula.

A \textit{course-prerequisite network} (CPN) is a directed graph where nodes represent courses offered at a university and directed links represent the prerequisite relationships between them. Namely, a directed link from node $i$ to node $j$ means that the course $i$ serves as a prerequisite for the course $j$. A CPN is a structural model of the university's academic curriculum, which represents the flow of knowledge in the curriculum and helps visualize, analyze, and optimize the complex system of courses that forms the core of the university's educational mission.  

In recent years, CPNs have attracted a lot of attention from various research groups due to their important role in quantifying and understanding academic curricula. For example, CPNs have been used for detecting critical courses~\cite{Slim2014a}, improving advising and curriculum reform~\cite{Aldrich2015}, studying the distribution of graduation time~\cite{Molontay2020}, and guiding curriculum design~\cite{Blas2021}. A more recent work~\cite{Stavrinides2023} proposed a general network-science-based framework for the analysis of CPNs, which allows users to identify important courses using classical centrality measures, describe the hierarchical structure of a CPN using the concept of topological stratification, quantify the strength of knowledge flow between different university divisions, and identify the most intradependent, influential, and interdisciplinary areas of study. This analysis was demonstrated using a network of courses taught at the California Institute of Technology. The CPNs of five Midwestern public universities were thoroughly analyzed in~\cite{Yang2024}, where a new graph theoretic measure of node importance, which is tailored to CPNs and quantifies the criticality of introductory courses, was introduced. 

To our knowledge, all CPN analyses conducted so far have primarily focused on either measuring the importance of \textit{individual nodes} within a CPN (using classical or new CPN-tailored centrality measures) or describing the \textit{internal organization} of a CPN (using various network structures, such as community structure and topological stratification). In other words, the analyses have focused only on the micro- and meso-scale properties of CPNs, and no macro-scale measures for an entire CPN have been defined (except for the basic notion of the size of the network).  

To fill this gap, in this paper, we propose and study three new \textit{global measures} of a whole CPN: \textit{breadth} and \textit{depth}, which quantify how wide and how deep the span of knowledge provided by the academic curriculum represented by the CPN is, and \textit{flux}, which quantifies the average amount of knowledge flow through the CPN. To illustrate these new measures, we use them to compare synthetic CPNs generated by two different network models, as well as CPNs of three universities: the Cyprus University of Technology (CUT), the California Institute of Technology (CIT), and Johns Hopkins University (JHU). The real CPN data analyzed in this paper are publicly available in the GitHub repository~\cite{GitHub}.

To define the breadth, depth, and flux of a CPN (Section~\ref{sec:New_CPN_Measures}), we first need to pre-process the network data via transitive reduction (Section~\ref{sec:TransitiveReduction}), discuss the difference between directed acyclic graphs and directed ordered graphs (Section~\ref{sec:DAGDOG}), and review the concept of topological stratification (Section~\ref{sec:Topological_Stratification}) on which the new CPN measures are based.

\section{Transitive Reduction of a CPN}
\label{sec:TransitiveReduction}

Similar to many other complex networks (transportation networks, citation networks, food webs, and the world wide web), a course-prerequisite network is a directed graph consisting of nodes connected by directed links. In a CPN, nodes represent courses that are offered at the university. A directed link $i\rightarrow j$ from node $i$ to node $j$ means that course $i$ serves as a \textit{prerequisite} for course $j$. In this case, we will also say that course $j$ is a \textit{postrequisite} for course $i$. For example, a course on differential geometry (DG) often serves as a prerequisite for a course on general relativity (GR). Here, DG is a prerequisite for GR, and GR is a postrequisite for DG.

Consider a directed graph containing three nodes $i$, $j$, and $k$ and three directed links $i\rightarrow j$, $j\rightarrow k$, and $i\rightarrow k$. In many applications (for instance, in the context of transportation networks or the world wide web), the link $i\rightarrow k$ can be interpreted as a ``shortcut'' that allows one to go from $i$ to $k$ without visiting $j$. In the context of CPNs, however,  $i\rightarrow k$ is a \textit{redundant} link: the fact that $i$ is a prerequisite for $k$ follows from the facts that $j$ is a prerequisite for $k$ (thanks to link $j\rightarrow k$) and $i$ is a prerequisite for $j$ (thanks to link $i\rightarrow j$). Given links $i\rightarrow j$ and $j\rightarrow k$, the link $i\rightarrow k$ does not provide any conceptually new information and its removal does not change the \textit{function} of the CPN.

\begin{figure}[h]
	\centering
	\includegraphics[width = 0.9\linewidth]{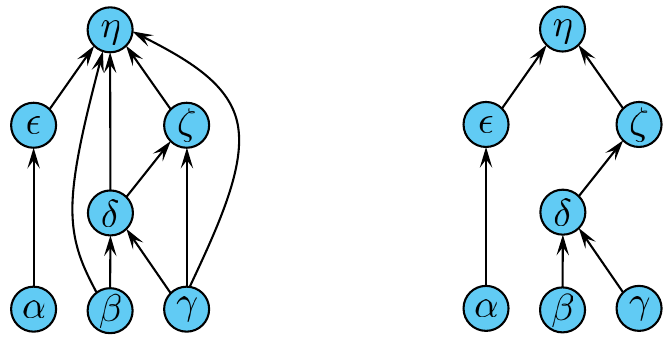}
	\caption{Left: A small directed acyclic graph $\mathcal{G}$. Right: The transitive reduction $\mathcal{G}^{tr}$ of $\mathcal{G}$. The links $\beta\rightarrow\eta$, $\delta\rightarrow\eta$, $\gamma\rightarrow\zeta$, and  $\gamma\rightarrow\eta$ are removed as redundant.}
	\label{fig:1}
\end{figure}

A \textit{transitive reduction} of a directed graph, introduced in \cite{Aho1972}, can be viewed as a mechanism for removing redundant links in a CPN. By definition, any CPN is a directed \textit{acyclic} graph (DAG), that is, a directed graph without cycles --- closed paths that start and end at the same node and follow links only in their forward direction. The transitive reduction of a directed acyclic graph $\mathcal{G}$ is a subgraph $\mathcal{G}^{tr}\subset\mathcal{G}$, which has the same nodes and is obtained from $\mathcal{G}$ by removing all links $i\rightarrow k$ such that there exists a longer directed path $i\rightarrow\ldots\rightarrow k$ from $i$ to $k$. The transitive reduction $\mathcal{G}^{tr}$ is uniquely defined, and it is the smallest (with respect to the number of links) subgraph of $\mathcal{G}$ that has the same reachability relation: there exists a directed path from node $i$ to node $j$ in $\mathcal{G}$ if and only if there exists a directed path from node $i$ to node $j$ in $\mathcal{G}^{tr}$. Figure~\ref{fig:1} shows a small DAG and its transitive reduction.

Since removing redundant links does not change a CPN conceptually and does not alter its function, any global measure of a CPN must be invariant with respect to transitive reduction. That is, if for any meaningful CPN measure $m$, we must have $m(\mathcal{G})=m(\mathcal{G}^{tr})$. In what follows, we will assume that instead of the original ``raw'' CPN, we work with its transitive reduction. Computing the transitive reduction can be viewed as a pre-processing step, as a ``cleaning'' of the network data.

\section{Dags and Dogs}
\label{sec:DAGDOG}

Directed acyclic graphs (DAGs) serve as models for many natural and man-made networks, such as citation networks~\cite{Radicchi2012}, where nodes represent documents (\eg, academic papers) and directed links represent citations between them, and food webs~\cite{Dunne2002}, where nodes represent species in an ecosystem and directed links represent predator-prey relationships. In many real networks, the acyclic structure is induced by a \textit{natural topological ordering} of the network nodes. Recall that a topological ordering of a directed graph with $n$ nodes is a total ordering of its nodes, $i_1<i_2<\ldots<i_n$, such that for each link $i_k\rightarrow i_m$ from node $i_k$ to node $i_m$, we have $i_k<i_m$, \ie, $i_k$ appears before $i_m$ in the ordering. A directed graph admits a topological ordering of its nodes if and only if it is a DAG, and the topological ordering is unique if and only if the graph has a Hamiltonian path (a directed path that visits all nodes exactly once)~\cite{Sedgewick}. 

Many networks are acyclic because the nature of the network imposes a natural topological ordering on its nodes. For example, in citation networks, papers can only cite other papers that have already been published: all links, therefore, must point backward in time. Thus, the publication time induces a natural topological ordering of the papers: $i<j$ if and only if paper $i$ was published after paper $j$. This time-induced ordering  is the fundamental reason why citation networks are acyclic. 

Following~\cite{Karrer2009}, we will refer to a DAG with a fixed topological ordering as a \textit{directed ordered graph} (DOG). Every DOG is clearly a DAG, and every DAG can be formally turned into a DOG by fixing one of its topological orderings. Unless a DAG has a Hamiltonian path, it has more than one topological ordering. For example, $\alpha<\beta<\gamma<\delta<\epsilon<\zeta<\eta$ and $\beta<\gamma<\delta<\alpha<\zeta<\epsilon<\eta$ are two orderings of the DAG in Fig.~\ref{fig:1}. Fixing an arbitrary topological ordering of a DAG, however, introduces an artificial (non-natural) \textit{total} order on the DAG's nodes, which are only \textit{partially} ordered by the DAG's links. 

Any real CPN is fundamentally a DAG but not a DOG, since it does not have a natural or canonical way to topologically order its nodes. In the next section, we describe a topological stratification~\cite{Stavrinides2023} of partially ordered CPN nodes, which is a generalization of topological ordering and the DAG structure on which all three proposed CPN measures are based.

\section{Topological Stratification}
\label{sec:Topological_Stratification}

\begin{figure}[h]
	\centering
	\includegraphics[width = 0.9\linewidth]{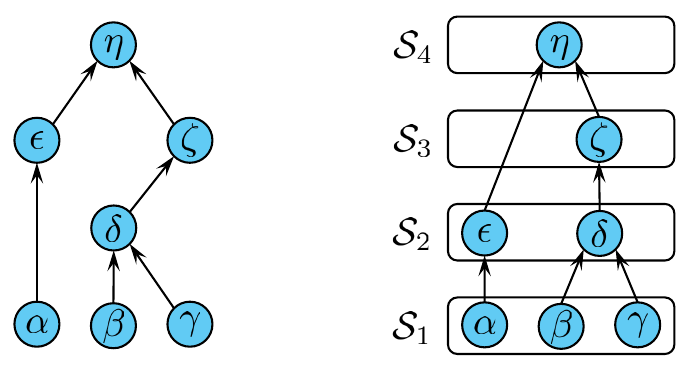}
	\caption{Left: A small course-prerequisite network. Right: Its topological stratification, consisting of four strata.}
	\label{fig:1.5}
\end{figure}

Let $\mathcal{G}=(\mathcal{V},\mathcal{E})$ be a CPN, where $\mathcal{V}$ and $\mathcal{E}$ are the sets of its nodes (vertices) and directed links (edges). A topological stratification of $\mathcal{G}$ is a partition of $\mathcal{V}$ into disjoint subsets, called \textit{strata}, of topologically equivalent nodes, which is defined as follows. The first stratum $\mathcal{S}_1\subset \mathcal{V}$ is the subset of nodes with zero in-degree (``sources''), that is, the set of courses with no prerequisites. The second stratum $\mathcal{S}_2\subset \mathcal{V}$ is obtained by first removing all nodes in $\mathcal{S}_1$ along with their outgoing links from the CPN, and then taking the nodes with zero in-degree in the remaining network. This process continues until all nodes are assigned to their strata. Let $T$ denote the total number of strata. As a result of this process, the CPN nodes are partitioned into $T$ disjoint strata,
\begin{equation}\label{eq:partition}
\mathcal{V}=\mathcal{S}_1\sqcup\mathcal{S}_2\sqcup\ldots\sqcup\mathcal{S}_T.
\end{equation}
This partition is illustrated in Fig.~\ref{fig:1.5} for a small CPN.

Each stratum $\mathcal{S}_t$ consists of \textit{topologically equivalent nodes}, \ie, nodes that are not connected by a directed path (courses that are not direct or indirect prerequisites or postrequisites of each other). Indeed, if $u\in \mathcal{S}_t$ and $u\rightarrow v$ is a directed link, then, after removing the first $(t-1)$ strata, the node $v$ has in-degree at least one, since $u$ is still in the network. This means that $v$ belongs to one of the higher strata: $v\in \mathcal{S}_{>t}=\mathcal{S}_{t+1}\sqcup\ldots\sqcup\mathcal{S}_T$. 

Within each stratum nodes are unordered, but the strata themselves are ordered. We refer to this ordered partition
\begin{equation}
	S(\mathcal{G})=\{\mathcal{S}_1,\ldots,\mathcal{S}_T\}
\end{equation}
as the \textit{topological stratification} of CPN $\mathcal{G}$. Note that the topological stratification is invariant with respect to the transitive reduction, namely, $S(\mathcal{\mathcal{G}}^{tr})=S(\mathcal{\mathcal{G}})$. This follows directly from the property established above that if $\mathcal{S}_t\ni u\rightarrow v$, then $v\in \mathcal{S}_{>t}$.

The topological stratification is a meso-scale structure of a CPN or, more generally, of a DAG. Other meso-structures often used for studying complex networks include the community structure~\cite{GirvanNewman2002, Porter2009, Fortunato2010}, core-periphery structure~\cite{Holme2005,Csermely2013,Rombach2014}, and $k$-core decomposition~\cite{Alvarez2005}. A common feature of all these constructions is that they partition the set of network nodes $\mathcal{V}$ into a collection of subsets according to a certain criterion. For the structures cited above, this criterion measures the density and sparsity of links within and between subsets of nodes or quantifies the ``coreness'' of the nodes. For the topological stratification, which is defined only for DAGs, this criterion is the topological equivalence of nodes, \ie, the absence of directed paths between nodes within the same subset (stratum). In the context of CPNs, the topological stratification  produces subsets $\mathcal{S}_1,\ldots,\mathcal{S}_T$ of courses of approximately the same level of difficulty, with more advanced courses lying in higher strata.

The topological stratification not only induces a \textit{hierarchy} of the CPN nodes with respect to the level of course advancement, but also helps describe the organization of the CPN links. Let $\mathcal{N}_v^-$ be the set of all prerequisites of node $v$, that is, the set of all nodes $u$ for which there is a link from $u$ to $v$. Similarly, let $\mathcal{N}_v^+$ be the set of all postrequisites of node $v$, that is, the set of all nodes $u$ for which there is a link from $v$ to $u$. Then $\mathcal{N}_v=\mathcal{N}_v^-\sqcup\mathcal{N}_v^+$ is the set of all ``neighbors'' of node $v$. The number of nodes in $\mathcal{N}_v^-$, $\mathcal{N}_v^+$, and $\mathcal{N}_v$ is, respectively, the in-, out-, and total degree of node $v$. Obviously,
\begin{equation}
	u\in\mathcal{N}_v^- \iff v\in\mathcal{N}_v^+.
\end{equation}

For any node $v$ in stratum $\mathcal{S}_t$, all its incoming links come from nodes lying in strictly lower strata, that is,
\begin{equation}
	\forall v\in\mathcal{S}_t, \hspace{3mm} \mathcal{N}_v^-\subset\mathcal{S}_{<t}=\sqcup_{i=1}^{t-1}\mathcal{S}_i.
\end{equation}
Indeed, by definition of $\mathcal{S}_t$, if we remove all nodes in $\mathcal{S}_1,\ldots,\mathcal{S}_{t-1}$ together with their outgoing links, then any $v\in\mathcal{S}_t$ will have zero in-degree. This means precisely that any incoming link of $v$ comes from a node lying in a lower stratum $\mathcal{S}_i$, where $i<t$.

Moreover, any node $v$ in stratum $\mathcal{S}_t$ must have at least one incoming link from a node lying in the strictly previous stratum:
\begin{equation}\label{eq:St-St+1}
\forall v\in\mathcal{S}_t, \hspace{3mm} \mathcal{N}_v^-\cap\mathcal{S}_{t-1}\neq \emptyset.
\end{equation}
Indeed, if this is not true and $\mathcal{N}_v^-\cap\mathcal{S}_{t-1}= \emptyset$, then $v$ will have zero in-degree after removing all nodes in strata $\mathcal{S}_1,\ldots,\mathcal{S}_{t-2}$ along with their outgoing links, which contradicts the assumption that $v \in \mathcal{S}_t$, as it would instead belong to $\mathcal{S}_{t-1}$ or an even lower stratum.

Next, as was already discussed above, for any node $v$ in stratum $\mathcal{S}_t$, all its outgoing links point to nodes lying in strictly higher strata, that is, 
\begin{equation}
	\forall v\in\mathcal{S}_t, \hspace{3mm} \mathcal{N}_v^+\subset\mathcal{S}_{>t}=\sqcup_{i=t+1}^{T}\mathcal{S}_i.
\end{equation}
However, there is no analog of property (\ref{eq:St-St+1}) for $\mathcal{N}_v^+$. Namely, a node $v\in\mathcal{S}_t$ may have no postrequisites in $\mathcal{S}_{t+1}$. In other words, if $v\in\mathcal{S}_t$, then the set $\mathcal{N}_v^-\cap\mathcal{S}_{t-1}$ may be empty.

\begin{figure}[h]
	\centering
	\includegraphics[width = 0.9\linewidth]{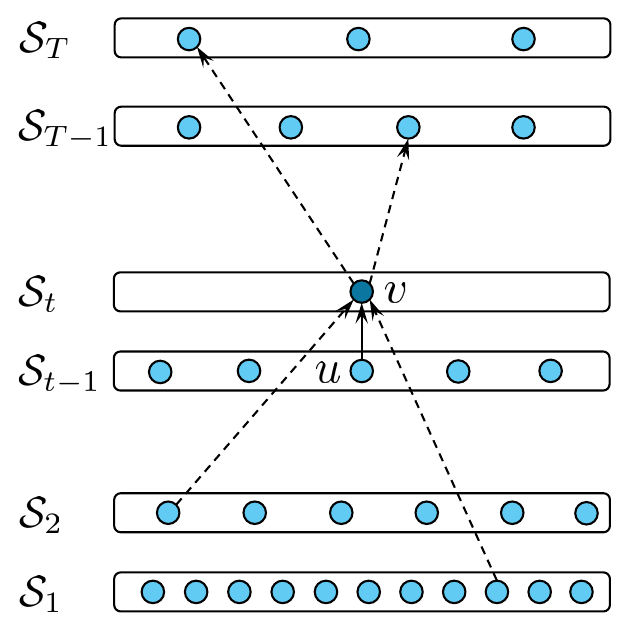}
	\caption{Every node $v\in\mathcal{S}_t$ has at least one prerequisite from a node  $u\in\mathcal{S}_{t-1}$ (a solid link that must exist), and all its prerequisites and postrequisites are in lower and higher strata, respectively (potential dashed links that may or may not exist).}
	\label{fig:2}
\end{figure}

To sum up, every CPN node lies in a single stratum of topologically equivalent nodes, all its prerequisites lie in lower strata, all its postrequisites lie in higher strata, and it has at least one prerequisite in the directly preceding stratum. The topological stratification of a CPN is schematically illustrated in Fig.~\ref{fig:2}.

This meso-scale structure serves as a basis for defining the proposed global CPN measures. 

\section{New CPN measures}
\label{sec:New_CPN_Measures}
The key quantitative characteristics of education are its \textit{breadth} and \textit{depth}, which describe, respectively, how wide the scope of knowledge provided by education is and how deep and advanced it is. We can define the breadth and depth of a CPN that represents an academic curriculum used for education as follows. 

\subsection{Breadth}
\label{sec:Breadth}

Let $\mathcal{G}$ be a CPN, $S(\mathcal{G})=\{\mathcal{S}_1,\ldots,\mathcal{S}_T\}$ be its topological stratification, and $n_t=|\mathcal{S}_t|$ be the number of nodes in stratum $\mathcal{S}_t$. It is natural to define the \textit{breadth} of $\mathcal{G}$ as the \textit{average stratum size}: 
\begin{equation}\label{eq:breadth}
	B(\mathcal{G})=\frac{1}{T}\sum_{t=1}^T|\mathcal{S}_t|=\frac{1}{T}\sum_{t=1}^Tn_t=\frac{n}{T},
\end{equation}
where $n$ is the total number of nodes in $\mathcal{G}$. This simple definition implies two  intuitively expected properties of the breadth: a) for two CPNs with the same number of strata, the one with more courses has greater breadth, and b) for two CPNs with the same number of courses, the one with a smaller number of strata has greater breadth. Note also that since the topological stratification is invariant with respect to 
the transitive reduction, so is the breadth: $B(\mathcal{G}^{tr})=B(\mathcal{G})$.

The number of strata $T$ has a useful interpretation: $T$ is the number of nodes in a longest directed path in $\mathcal{G}$ (there may be several different longest directed paths in $\mathcal{G}$, but the number of nodes in any such path is the same).

Indeed, let $L$ be the number of nodes in a longest directed path. On the one hand, $L\leq T$ since if $L$ were larger than $T$, then there would be at least two nodes in the same stratum connected by a link, which is impossible (since there are no links between nodes lying in the same stratum). On the other hand, property (\ref{eq:St-St+1}), namely the presence of a solid link in Fig.~\ref{fig:2}, allows us to construct a directed path of length $T$ as follows. Consider any node $\omega\in\mathcal{S}_T$ in the last stratum. It necessarily has a prerequisite in the previous stratum $\omega\in\mathcal{S}_{T-1}$, which has a prerequisite in $\mathcal{S}_{T-2}$, and so one, until a zero in-degree prerequisite $\alpha\in\mathcal{S}_1$ in the first stratum is reached. The corresponding directed path $\alpha\rightarrow\ldots\rightarrow\omega$ has $T$ nodes. Therefore, the number of nodes in a longest path $L\geq T$. Combining $L\leq T$ and $L\geq T$ leaves us with the only possibility: $L=T$. 

So, the breadth $B(\mathcal{G})$ of a CPN $\mathcal{G}$ is the ratio of the total number of nodes $n$ of the CPN to the number of nodes $T$ in its longest directed path. The larger the $n$, the broader the network; the longer a longest path, the narrower the CPN.

\subsection{Depth}
\label{sec:Depth}

Intuitively, the depth of a CPN should be defined in terms of the lengths of directed paths in the network, and it should be negatively correlated with the breadth: the longer the paths, the deeper the network is. This idea can be formalized as follows. 

Let $\Omega\subset \mathcal{V}$ be the set of all nodes with zero out-degree. Nodes $\omega\in\Omega$ represent the most advanced  (``deepest'') courses students can possibly take by traversing $\mathcal{G}$. We define the \textit{depth} of $\mathcal{G}$ as the \textit{average depth of nodes in} $\Omega$:
\begin{equation}
	D(\mathcal{G})=\frac{1}{|\Omega|}\sum_{\omega\in\Omega} d(\omega),
\end{equation}
where $|\Omega|$ is the number of nodes in $\Omega$ and $d(\omega)$ is the depth of node $\omega$, a quantity which is still to be defined. 

Let $\Omega_t=\Omega\cap\mathcal{S}_t$ be the subset of nodes with zero out-degree lying in the stratum $\mathcal{S}_t$, and 
\begin{equation}
	\Omega=\Omega_1\sqcup\Omega_2\sqcup\ldots\sqcup\Omega_T
\end{equation}
be the partition of $\Omega$ induced by the partition of $\mathcal{V}$ (\ref{eq:partition}). Then the CPN depth can be rewritten as follows:
\begin{equation}
D(\mathcal{G})=\frac{1}{|\Omega|}\sum_{t=1}^T\sum_{\omega\in\Omega_t} d(\omega).
\end{equation}

\begin{figure}[h]
	\centering
	\includegraphics[width = 0.9\linewidth]{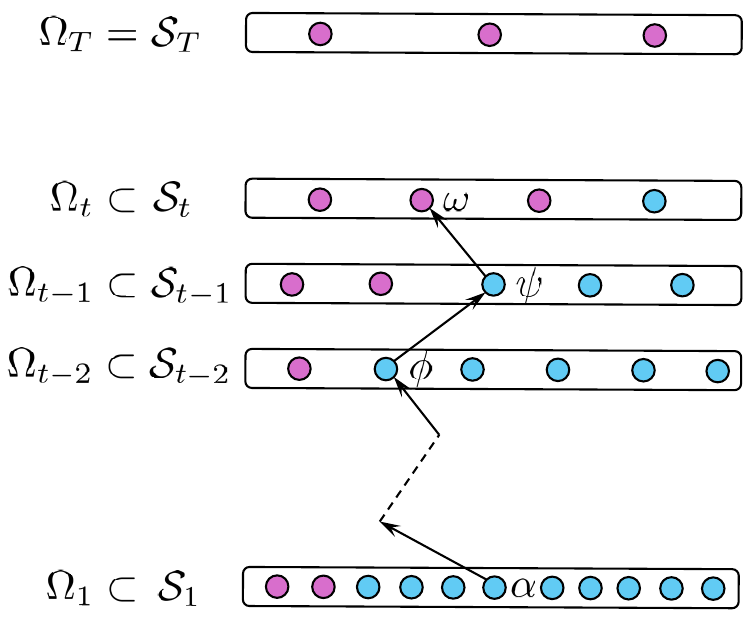}
	\caption{Every stratum $\mathcal{S}_t$ is partitioned into the subset $\Omega_t$ of zero out-degree (purple) nodes and the subset of positive out-degree (blue) nodes. The depth of $\omega\in\Omega_t$ is the number of nodes in the longest path $\alpha\rightarrow\ldots\rightarrow\phi\rightarrow\psi\rightarrow\omega$. The depth of the CPN is the average depth of nodes in $\Omega$.}
	\label{fig:3}
\end{figure}

Consider a node $w\in\Omega_t$. Since $\omega\in\mathcal{S}_t$, according to the property (\ref{eq:St-St+1}), it has at least one prerequisite $\psi\in\mathcal{S}_{t-1}$ in the previous stratum, which, in turn, has at least one prerequisite $\phi\in\mathcal{S}_{t-2}$, and so on until a course $\alpha\in\mathcal{S}_{1}$ in the first stratum is reached. Taking a sequence of courses $\alpha\rightarrow\ldots\rightarrow\phi\rightarrow\psi$ is a necessary condition for taking $\omega$ (but not necessarily sufficient since $\omega$ may have other direct and indirect prerequisites). The path $\alpha\rightarrow\ldots\rightarrow\phi\rightarrow\psi\rightarrow\omega$ is the \textit{longest directed path} in the CPN that ends in $\omega$, see Fig.~\ref{fig:3}. 
We define the depth of $\omega\in\Omega_t$ as the number of nodes in that path, that is:
\begin{equation}
	d(\omega)=t, \hspace{3mm}\omega\in\Omega_t.
\end{equation}

The depth of the CPN $\mathcal{G}$ is, therefore,
\begin{equation}\label{eq:depth}
D(\mathcal{G})=\frac{1}{|\Omega|}\sum_{t=1}^T\sum_{\omega\in\Omega_t} t=\frac{1}{|\Omega|}\sum_{t=1}^T t|\Omega_t|,
\end{equation}
where $|\Omega_t|$ is the number of nodes in $\Omega_t$.

This definition of the CPN depth is invariant with respect to the transitive reduction, $D(\mathcal{G}^{tr})=D(\mathcal{G})$, and it has a simple probabilistic interpretation. Indeed, 
\begin{equation}
D(\mathcal{G})=	\sum_{t=1}^T t\frac{|\Omega_t|}{|\Omega|}=\sum_{t=1}^T t\mathbb{P}(\omega^*\in\Omega_t)=\mathbb{E}[t^*],
\end{equation}
where $\omega^*$ is a random node chosen uniformly from $\Omega$, and $t^*$ is the random index of the stratum to which $\omega^*$ belongs, that is, $\omega^*\in\mathcal{S}_{t^*}$. The depth of a CPN is then the expected value of the index of the stratum that contains a randomly chosen node with zero out-degree. 

The definition of the CPN breadth (\ref{eq:breadth}) can also be interpreted probabilistically:
\begin{equation}
	B(\mathcal{G})=\frac{1}{T}\sum_{t=1}^T|\mathcal{S}_t|=\sum_{t=1}^T|\mathcal{S}_t|\mathbb{P}(t^\star=t)=\mathbb{E}[|\mathcal{S}_{t^\star}|],
\end{equation}
where $t^\star$ is a random index chosen uniformly from the set $\{1,\ldots,T\}$, and $|\mathcal{S}_{t^\star}|$ is the size of stratum $\mathcal{S}_{t^\star}$. So, the breadth of a CPN is then the expected size of a randomly chosen stratum (which is the average stratum size).

\subsection{Flux}
\label{sec:Flux}
\begin{figure*}[t!]
	\centering
	\includegraphics[width = \linewidth]{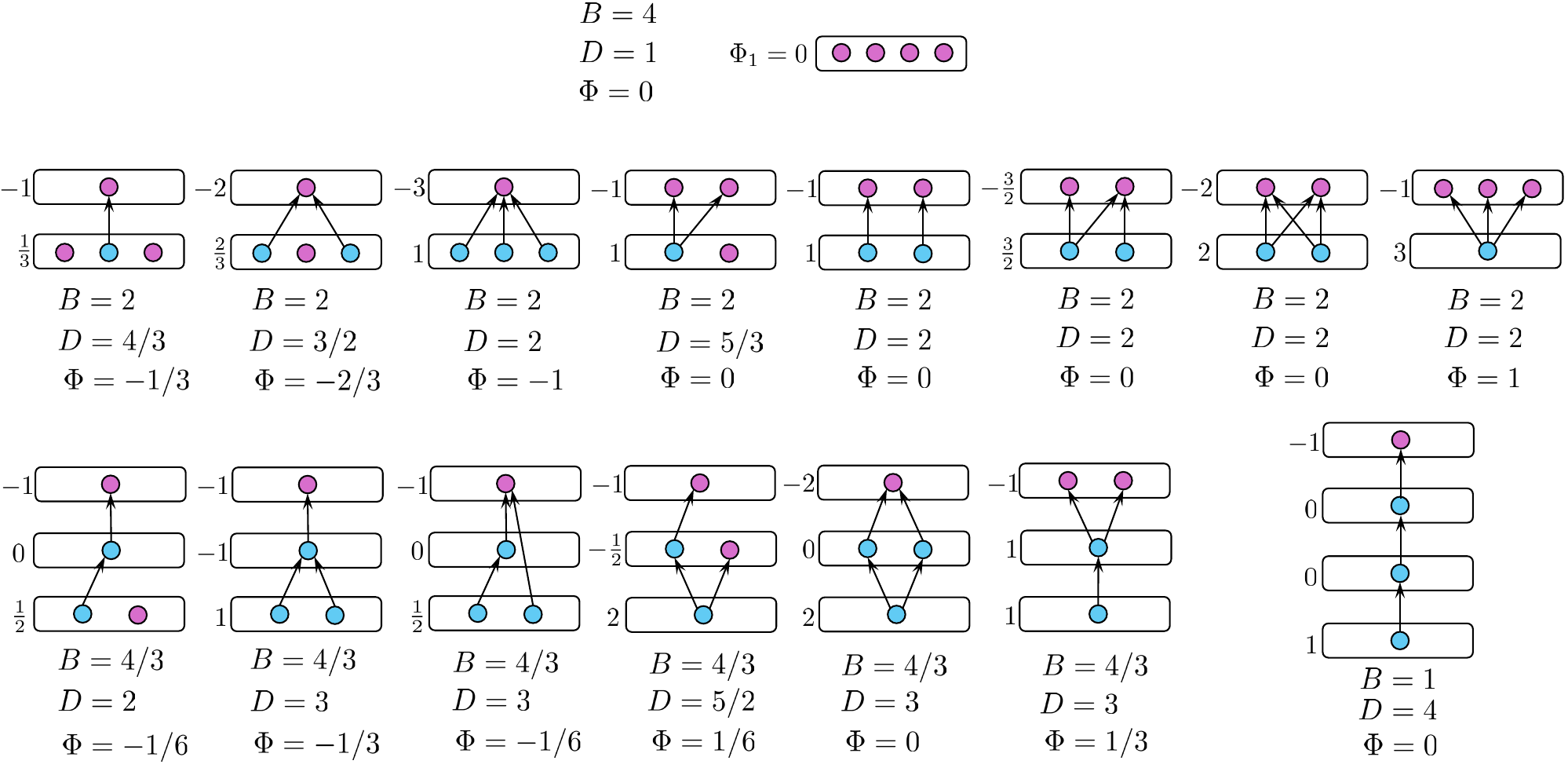}
	\caption{All 16 different CPNs with 4 nodes and their breadths (B), depths (D), and fluxes ($\Phi$), computed via (\ref{eq:breadth}), (\ref{eq:depth}), and (\ref{eq:flux}), respectively. The purple nodes indicate the nodes with zero out-degree, which are used for computing the CPN depth. The numbers next to strata are the values of fluxes $\Phi_t$ through those strata, as defined in (\ref{eq:flux_t}).}
	\label{fig:4}
\end{figure*}

The last proposed CPN measure is the concept of \textit{flux}, which is inspired and motivated by a similar concept introduced by Karrer and Newman in~\cite{Karrer2009}. In that paper, however, the flux is defined as a property of the gap between two successive nodes of a DOG (a directed acyclic graph with a fixed total order of nodes, see Section~\ref{sec:DAGDOG}). Since a CPN is a DAG, but not a DOG, and it does not have a canonical or natural way to order its nodes, we define the flux at the level of strata, not at the level of gaps between successive nodes.

Conceptually, a CPN represents the \textit{flow of knowledge} in an academic curriculum, where knowledge ``flows'' along directed paths in the CPN from lower strata containing more introductory courses to higher strata containing more advanced courses. In the physical sciences that study transport phenomena, the flux $\Phi(t)$ at time $t$ is defined as the rate of flow of a certain quantity per unit area,
\begin{equation}\label{eq:flux_classal}
	\Phi(t)=\frac{\dot{q}(t)}{A}=\frac{1}{A}\lim_{\Delta t\rightarrow0}\frac{q(t+\Delta t)-q(t)}{\Delta t}
\end{equation}
where $q(t)$ is the quantity that flows and $A$ is the area of the surface through which the quantity flows.
A discrete analog of (\ref{eq:flux_classal}), in the context of CPNs, is the \textit{flux $\Phi_t$ through stratum $\mathcal{S}_t$}, which we define as follows:
\begin{equation}\label{eq:flux_t}
\Phi_t=\frac{L_t^{t+1}-L_{t-1}^{t}}{n_t},
\end{equation}
where $n_t=|\mathcal{S}_t|$ is the number of nodes in stratum  $\mathcal{S}_t$, $L_{t-1}^{t}$ is the number of incoming links from  $\mathcal{S}_{t-1}$ to  $\mathcal{S}_t$, and $L_t^{t+1}$ is the number of outgoing links from $\mathcal{S}_t$ to $\mathcal{S}_{t+1}$. Here, $n_t$ quantifies the size of $\mathcal{S}_t$ and plays the role of the area $A$ in (\ref{eq:flux_classal}). The number of links $L_{t-1}^{t}$ between two successive strata acts as the flowing quantity $q(t)$. 

Note that in the definition (\ref{eq:flux_t}) of the flux $\Phi_t$ through stratum $\mathcal{S}_t$, we intentionally do not count incoming links from $\mathcal{S}_{<t-1}=\mathcal{S}_1\sqcup\ldots\sqcup\mathcal{S}_{t-2}$ to $\mathcal{S}_t$ and outgoing links from $\mathcal{S}_t$ to $\mathcal{S}_{>t+1}=\mathcal{S}_{t+2}\sqcup\ldots\sqcup\mathcal{S}_{T}$. The reason is twofold. First, $\Phi_t$ is intended to be a ``local'' property of $\mathcal{S}_t$ that quantifies its interactions with the neighboring strata $\mathcal{S}_{t-1}$ and $\mathcal{S}_{t+1}$. Second, $\Phi_t$ must be invariant with respect to the transitive reduction. The definition (\ref{eq:flux_t}) has this property since links from  $\mathcal{S}_{t-1}$ to  $\mathcal{S}_t$ and from $\mathcal{S}_{t}$ to  $\mathcal{S}_{t+1}$ are present in both $\mathcal{G}$ and $\mathcal{G}^{tr}$ (these links are not redundant in the sense of Section~\ref{sec:TransitiveReduction}). Counting links from $\mathcal{S}_{<t-1}$ to $\mathcal{S}_t$ or from $\mathcal{S}_t$ to $\mathcal{S}_{>t+1}$, that is, defining the flux as $\Phi_t=(L_{t}^{t+1:T}-L_{1:t-1}^t)/n_t$, where $L_{t}^{t+1:T}$ and $L_{1:t-1}^t$ are the numbers of links, respectively, from $\mathcal{S}_t$ to $\mathcal{S}_{>t}$ and from $\mathcal{S}_{<t}$ to $\mathcal{S}_t$, will destroy the invariance. For example, removing a redundant link from $\mathcal{S}_{<t-1}$ to $\mathcal{S}_t$ will decrease $L_{1:t-1}^t$ and increase $\Phi_t$. Similarly, removing a redundant link from $\mathcal{S}_t$ to $\mathcal{S}_{>t+1}$ will decrease $L_{t}^{t+1:T}$ and decrease $\Phi_t$.

The flux $\Phi_t$ through stratum $\mathcal{S}_t$ can be expressed as the average of local fluxes through nodes of $\mathcal{S}_t$. Let $A$ be the adjacency matrix of the CPN $\mathcal{G}$, where $A_{ij}=1$ if there is a link $i\rightarrow j$ and $A_{ij}=0$ otherwise. Then 
\begin{equation}
\begin{split}
L_t^{t+1}&=\sum_{i\in\mathcal{S}_t}\sum_{j\in\mathcal{S}_{t+1}}A_{ij},\\ L_{t-1}^{t}&=\sum_{i\in\mathcal{S}_t}\sum_{k\in\mathcal{S}_{t-1}}A_{ki}.
\end{split}
\end{equation}
Therefore, 
\begin{equation}
\begin{split}
&\Phi_t=\frac{L_t^{t+1}-L_{t-1}^{t}}{n_t}\\
&=\frac{1}{n_t}\sum_{i\in\mathcal{S}_t}\left(\sum_{j\in\mathcal{S}_{t+1}}A_{ij}-\sum_{k\in\mathcal{S}_{t-1}}A_{ki}\right)=\frac{1}{n_t}\sum_{i\in\mathcal{S}_t}\phi_i,
\end{split}
\end{equation}
where
\begin{equation}
\phi_i=\sum_{j\in\mathcal{S}_{t+1}}A_{ij}-\sum_{k\in\mathcal{S}_{t-1}}A_{ki}
\end{equation}
is the \textit{local flux through node $i\in\mathcal{S}_t$}. 

The flux $\Phi_t$ quantifies the average amount of ``knowledge flow'' that is emitted from ($\Phi_t>0$) or absorbed by ($\Phi_t<0$) the nodes in stratum $\mathcal{S}_t$. The more courses in $\mathcal{S}_t$ collectively serve as prerequisites for courses in $\mathcal{S}_{t+1}$ compared to their role as postrequisites for courses in  $\mathcal{S}_{t-1}$, the larger the value of $\Phi_t$. Strata consisting of introductory courses tend to have positive flux, while those consisting of more advanced courses tend to have negative flux.

Finally, we define the \textit{flux of a CPN} as the average flux through its strata:
\begin{equation}\label{eq:flux}
	\Phi=\frac{1}{T}\sum_{t=1}^T\Phi_t.
\end{equation}
The CPN flux can be positive, negative, or zero, depending on the values of $\Phi_t$. For example, the flux through the first stratum is always positive, $\Phi_1>0$ (except in the extreme case where the CPN consists of a single stratum, in which case $\Phi=\Phi_1=0$), and the flux through the last stratum is always negative, $\Phi_T<0$. 

To illustrate the proposed CPN measures and provide the reader with a better intuitive understanding of these measures, Fig.~\ref{fig:4} shows all 16 different CPNs consisting of 4 nodes (after topological reduction and up to isomorphism), along with their breadths, depths, and fluxes. As intuitively expected, the breadth and depth are negatively correlated. The average values of depth for breadth $B=1, 4/3, 2, 4$ are, respectively, $D=4, 2.75, 1.8125, 1$.

\section{Two Random Graph CPN Models}
\label{sec:CPN_moidels}

In this paper, we compare the breadth, depth, and flux of three real CPNs --- the course networks of the Cyprus University of Technology (CUT), the California Institute of Technology (CIT), and Johns Hopkins University (JHU) --- as well as those of synthetic CPNs generated by two random graph models, which are roughly analogous to the classical Erd\H{o}s–R\'{e}nyi model~\cite{ER1959} and the standard configuration model~\cite{Bollobas1980} for undirected graphs. 

It is important to highlight that we consider these two random graph models not because they generate DAGs similar to real CPNs (as we will see, they do not). We leave the development of a good random graph model for CPNs to future research. Here, our goal is to compare the breadth, depth, and flux of the real CPNs of CUT, CIT, and JHU with those of other DAGs that have the same numbers of nodes and links or the same degree sequence. For this purpose, we use synthetic CPNs (model-generated DAGs) because real CPNs are surprisingly difficult to find, and to the best of our knowledge, there are no other public datasets containing entire university CPNs, except for our GitHub repository~\cite{GitHub}.

\subsection{The Erd\H{o}s–R\'{e}nyi CPN Model}
\label{sec:ER_model}

As discussed in Section~\ref{sec:DAGDOG}, any CPN is fundamentally a DAG (but not a DOG). A directed graph is acyclic if and only if its nodes can be topologically ordered. In other words, the nodes can be ordered (numerically labeled) such that the adjacency matrix $A$ of the graph with respect to that order is upper triangular. 

Let $U(n,m)$ be the set of all $n \times n$ strictly upper triangular binary matrices (whose elements are zeros and ones) with exactly $m$ ones. Let $A\in U(n,m)$ be a random $n \times n$ matrix with $m$ ones uniformly distributed above the diagonal. Algorithmically, $A$ can be generated by starting with the zero $n \times n$ matrix, choosing $m$ out of $n(n-1)/2$ elements above the diagonal via random sampling without replacement, and replacing them with ones. Matrix $A$ generated this way is sampled uniformly at random from $U(n,m)$. We say that the random graph $G$ with the adjacency matrix $A$ is generated according to the Erd\H{o}s–R\'{e}nyi CPN model, denoted
\begin{equation}
	G\sim \mathrm{ER}(n,m).
\end{equation}

The random graph $G$ is then sampled uniformly from the ensemble of all DOGs with $n$ nodes and $m$ links. It is not, however, uniformly distributed on the ensemble of all DAGs with $n$ nodes and $m$ links, because two different DOGs can be isomorphic to the same DAG after dropping the fixed topological ordering (numerical labels on nodes). To the best of our knowledge, no method of sampling uniformly from the ensemble of DAGs is  known. 

\subsection{The Karrer-Newman CPN Model}
\label{sec:KN_model}
The Karrer-Newman (KN) CPN model is based on the KN model for DOGs~\cite{Karrer2009,Karrer2009-2}, which is an analog of the standard configuration model for undirected graphs. 

Consider a DOG with $n$ nodes, \ie, a directed acyclic graph with a fixed topological ordering of the nodes, denoted by $1,\ldots,n$. Thus, the graph can have a link from node $i$ to node $j$ only if $i<j$. Let $k_i^{\mathrm{in}}$ and $k_i^{\mathrm{out}}$ be the in- and out-degree of node $i$, and let 
\begin{equation}
	(k^{\mathrm{in}},k^{\mathrm{out}})=(k^{\mathrm{in}}_1,k^{\mathrm{out}}_1),\ldots,(k^{\mathrm{in}}_n,k^{\mathrm{out}}_n)
\end{equation} 
be the corresponding degree sequence of the graph. 

\begin{figure}[h]
	\centering
	\includegraphics[width = 0.2\linewidth]{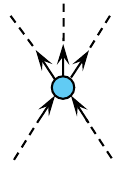}
	\caption{A node with 2 incoming and 3 outgoing stubs.}
	\label{fig:5}
\end{figure}

The original KN model takes the ordered degree sequence $(k^{\mathrm{in}},k^{\mathrm{out}})$ as its input and generates a random DOG with precisely this degree sequence as follows. It is convenient to represent the degree sequence as a collection of link ``stubs'' (half-links) pointing in and out of nodes in the appropriate numbers. As an illustration, Fig.~\ref{fig:5} shows a node with in-degree 2 and out-degree 3 represented by stubs. 

Now, given $n$ nodes with no links but with the appropriate number of incoming and outgoing stubs at each node, as described by the degree sequence $(k^{\mathrm{in}},k^{\mathrm{out}})$, we visit all nodes in order from 2 to $n$. For each node $i$, we randomly attach all its $k_i^{\mathrm{in}}$ incoming stubs to $k_i^{\mathrm{in}}$ outgoing stubs of previous nodes $1,\ldots,i-1$, chosen uniformly from the set of all such outgoing stubs that are currently unattached. This process allows multilinks (just as in the standard configuration model), but they usually constitute a negligible fraction of all links. It can be shown that this model generates a random DOG sample uniformly from the ensemble of all DOGs (possibly with multilinks) with the degree sequence $(k^{\mathrm{in}},k^{\mathrm{out}})$~\cite{Karrer2009,Karrer2009-2}. 

The KN CPN model is based on the original KN model for DOGs described above. It takes a real CPN $\mathcal{G}$ as its input and generates a random CPN $G$ with the same unordered degree sequence. Since any real CPN $\mathcal{G}$ is a DAG and has many possible topological orderings of its nodes, to use the original KN models, we need to first convert the DAG $\mathcal{G}$ into a DOG by selecting a specific ordering. Thus, given $\mathcal{G}$, a random CPN $G$ is generated as follows:
\begin{enumerate}
	\item Compute the topological stratification of $\mathcal{G}$,
	\begin{equation}
	S(\mathcal{G})=\{\mathcal{S}_1,\ldots,\mathcal{S}_T\}.
	\end{equation}
	\item For each stratum $\mathcal{S}_t$, select a random ordering of its nodes uniformly from the set of all $n_t!$ permutations of $n_t$ nodes. 
	\item Concatenate the orderings obtained in step 2 into a single ordering of nodes of $\mathcal{G}$, which, by construction, will be a topological ordering of $\mathcal{G}$. 
	\item Let $(k^{\mathrm{in}},k^{\mathrm{out}})$ be the degree sequence of $\mathcal{G}$ with respect to the topological ordering from step 3. 
	\item Generate a random graph $G$ using the original KN model with the degree sequence $(k^{\mathrm{in}},k^{\mathrm{out}})$.
	\item Replace all multilinks in $G$ (if any) with single links. 
\end{enumerate}

We say that the random graph $G$ produced this way is generated according to the Karrer-Newman CPN model, denoted
\begin{equation}
	G\sim \mathrm{KN}(\mathcal{G}).
\end{equation}
This notation highlights that generating a random CPN $G$ that mimics the degree properties of $\mathcal{G}$ requires the full network $\mathcal{G}$.

\section{Empirical and Simulation Results}
\label{sec:Simulation}

We consider three real CPNs corresponding to three very different universities: one European university, Cyprus University of Technology (CUT), and two US universities, California Institute of Technology (CIT) and Johns Hopkins University (JHU). CUT is the smallest and youngest university (established in 2004), located in Limassol, Cyprus. It comprises seven faculties and a language center. It was established to fill gaps in the Cyprus higher education system by offering degrees not provided by other institutions. CIT is a top-tier private research university, based in Pasadena, California, with a very strong emphasis on research in STEM fields. It has the highest number of Nobel laureates per capita (as of October 2024), is consistently ranked among the top universities in the world, and has more graduate than undergraduate students. JHU is the largest and oldest of the three considered universities (founded in 1876), located in Baltimore, Maryland. It is considered the first research university in the US. Compared to CIT, JHU is much larger, more diverse academically, offers a broader curriculum, and is renowned for its medical and public health research. 

Table~\ref{tab:1} summarizes the five global CPN measures for CUT, CIT, and JHU: the number of nodes $n$,  the number of links $m$ (after transitive reduction), breadth $B$, depth $D$, and flux $\Phi$. As intuitively expected, the larger the size $n$ of the network (\ie, the more courses are offered by the curriculum), the larger its breadth $B$. Also, as discussed in section~ \ref{sec:Depth}, the breadth and depth are negatively correlated: when new courses are added to the curriculum, they are not likely to increase the lengths of the longest paths.
\begin{table}[H]
	\centering
	\begin{tabular}{c|c|c|c|c|c}
		\textsc{{}} & $n$ & $m$ & $B$ & $D$ & $\Phi$  \\
		\hline
		CUT & 416 & 319 & 59.43 & 2.66 & -0.31\\
		CIT & 771 & 640 & 110.14 & 2.37 & -0.35\\
		JHU & 10291 & 3499 & 857.58 & 1.34 &-0.55
	\end{tabular}
	\caption{Three real CPNs and their five global measures.}
	\label{tab:1}
\end{table}
\begin{figure}[h]
	\centering
	\includegraphics[width = \linewidth]{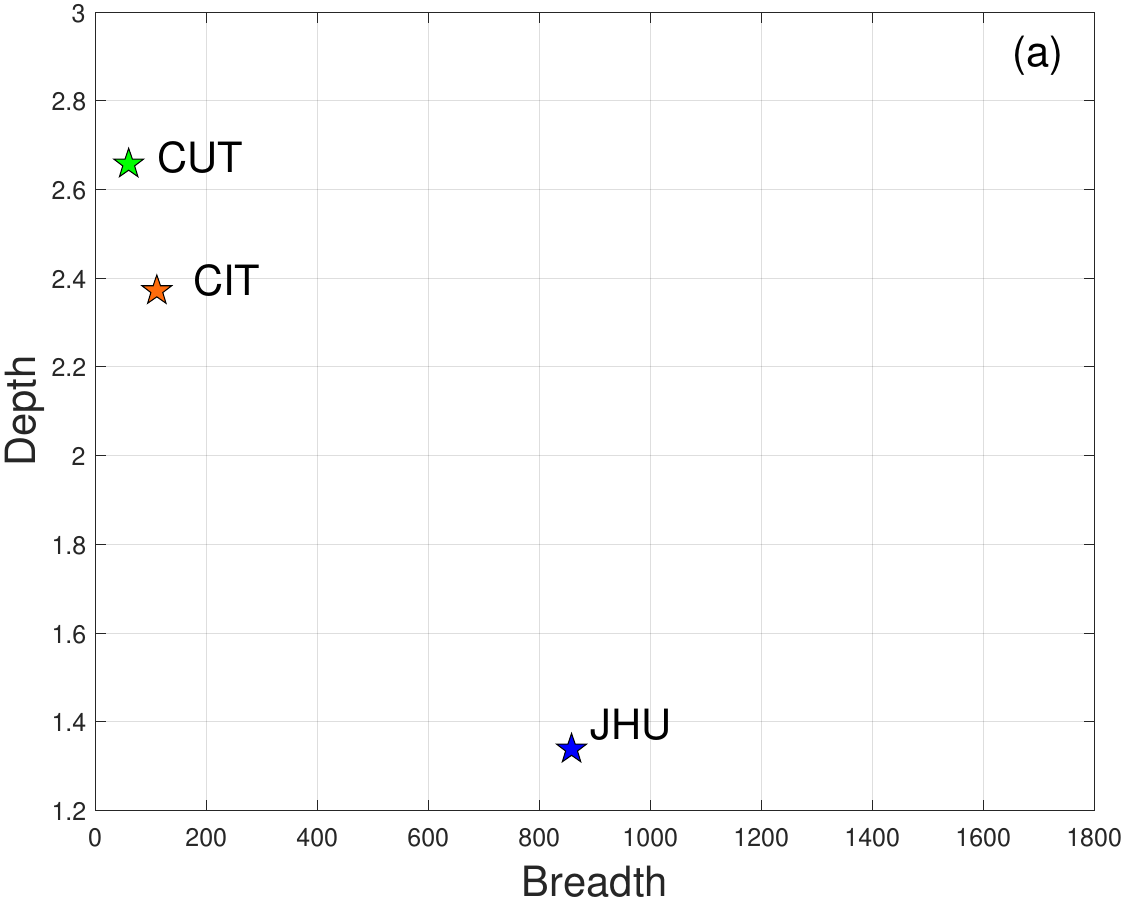}
	\includegraphics[width = \linewidth]{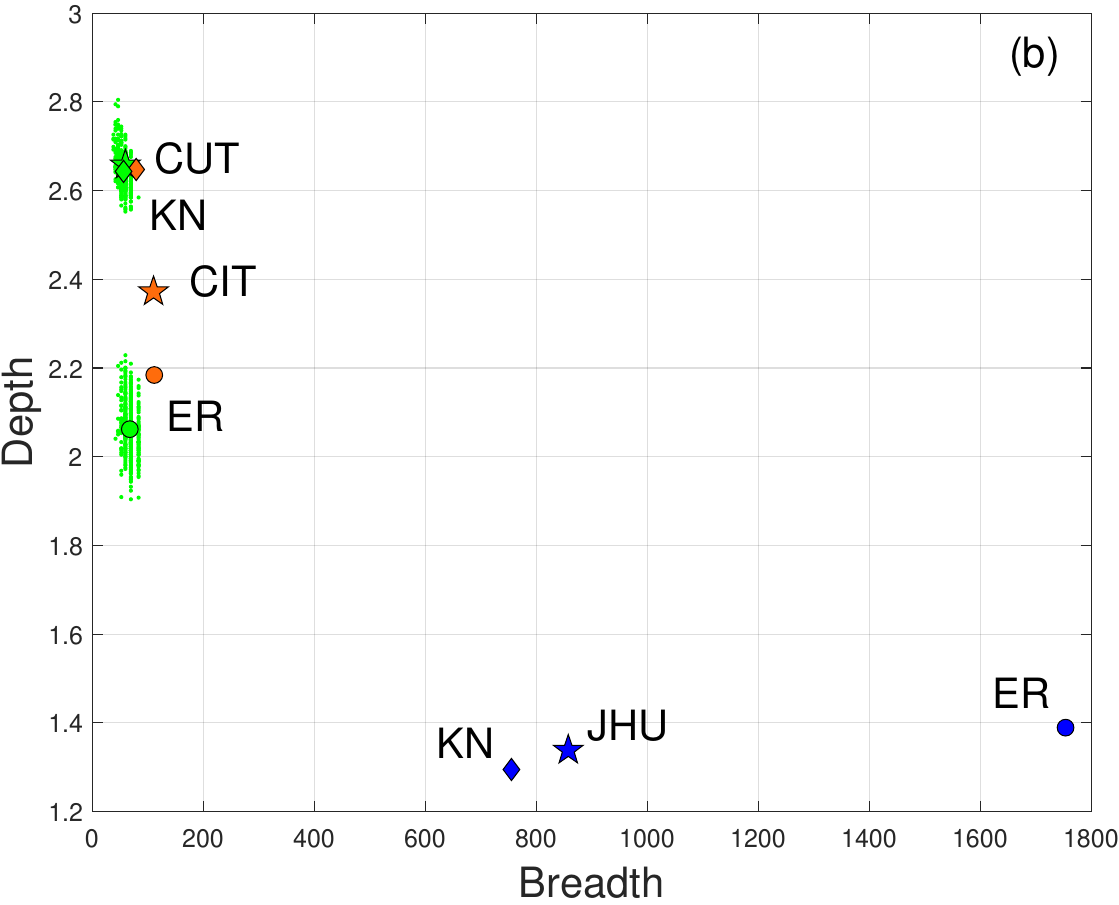}
	\caption{The depth versus breadth for CUT, CIT, JHU, and synthetic CPNs generated by the ER and KN models.}
	\label{fig:6}
\end{figure}
\begin{figure}[h]
	\centering
	\includegraphics[width = \linewidth]{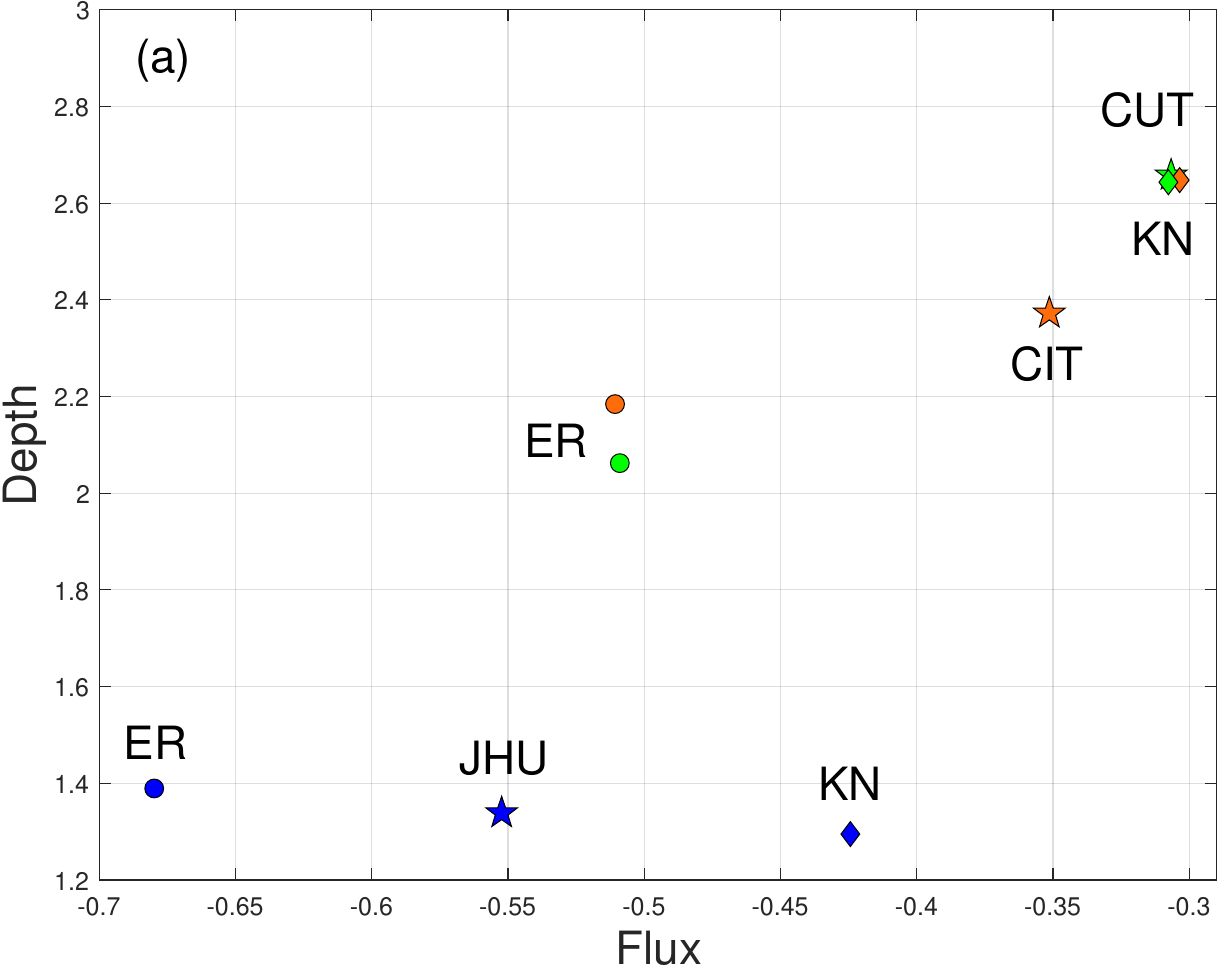}
	
	\vspace{1mm}
	\includegraphics[width = \linewidth]{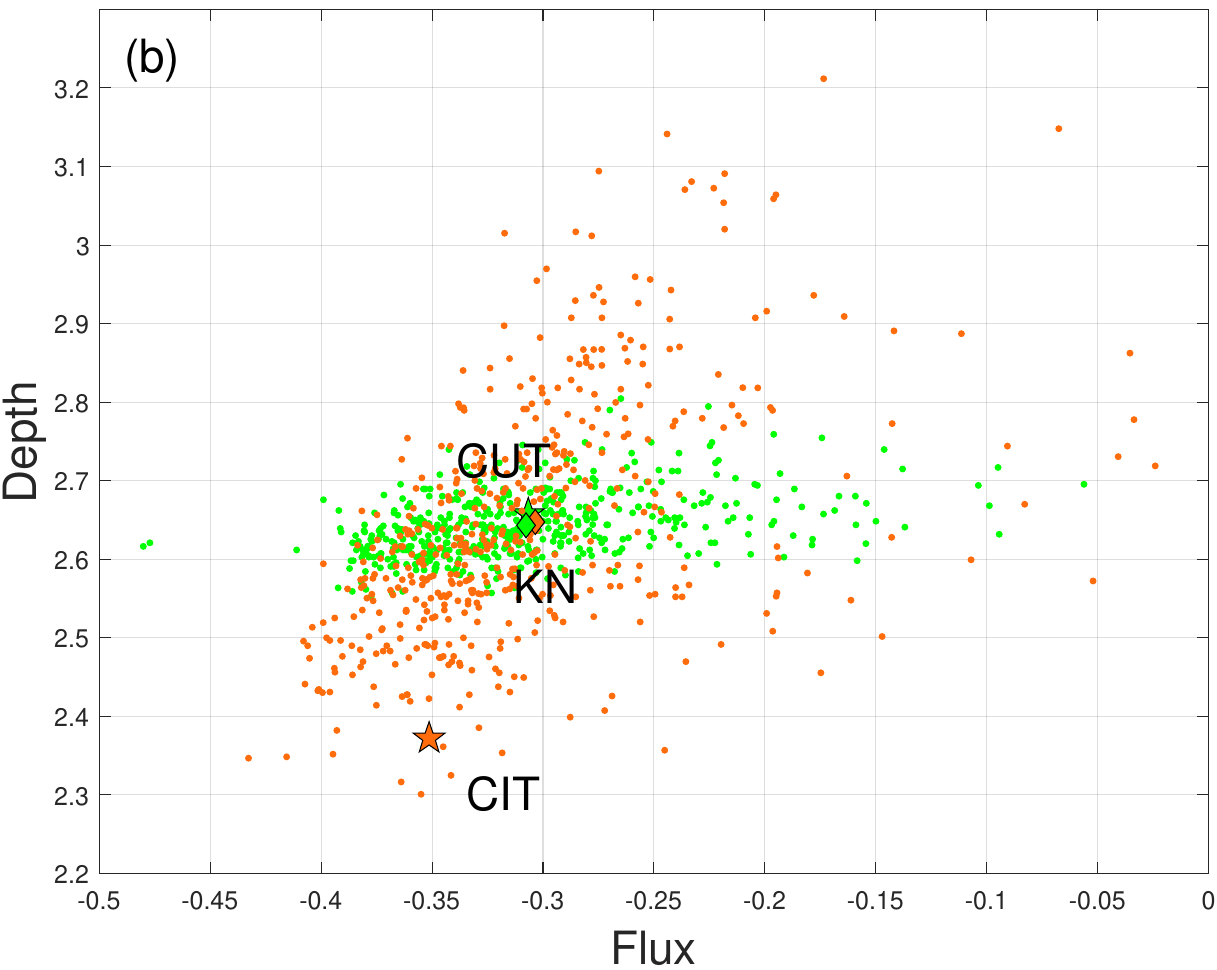}
	
	\vspace{.7mm}
	\caption{The depth versus flux for CUT, CIT, JHU, and synthetic CPNs generated by the KN model.}
	\label{fig:7}
\end{figure}

Figure~\ref{fig:6}(a) shows depth $D$ versus breadth $B$ for the three universities: CUT (green), CIT (orange), and JHU (blue). We compare these true values with synthetic ones obtained from $N=500$ random graphs generated according to both the Erd\H{o}s–R\'{e}nyi (ER) and Karrer-Newman (KN) CPN models and plot the average values of their depths and breadths in Fig.~\ref{fig:6}(b). The average values of ER- and KN-generated graphs are represented by circles and diamonds, respectively. While the ER model deviates substantially, the KN model, on average, accurately captures CUT's breadth and depth (the green diamond is close to the green star). The green scatter plots show the individual values generated by the ER and KN models for CUT, namely,  the values of depth and breadth for $G_1,\ldots,G_N\sim \mathrm{ER}(n_{\mathrm{CUT}},m_{\mathrm{CUT}})$ and $G_1,\ldots,G_N\sim \mathrm{KN}(\mathcal{G}_{\mathrm{CUT}})$. For CIT and JHU, however, both models fail to capture the real CPNs' breadth and depth, and the corresponding scatter plots are not shown.

In Figure~\ref{fig:7}(a), we plot depth $D$ versus flux $\Phi$ for CUT, CIT, and JHU (stars), as well as the average values of $D$ and $\Phi$ computed from $N=500$ CPNs generated by the ER (circles) and KN (diamonds) models. As before, the ER model is inaccurate in all three cases, and the KN model accurately captures, on average, only the depth and flux of CUT, the smallest of the three real CPNs. Figure~\ref{fig:7}(b) shows scatter plots of the individual values generated by the KN model for CUT (green) and Caltech (orange), namely, the values of depth and flux for $G_1,\ldots,G_N\sim \mathrm{KN}(\mathcal{G}_{\mathrm{CUT}})$ and $G_1,\ldots,G_N\sim \mathrm{KN}(\mathcal{G}_{\mathrm{CIT}})$.

\begin{figure}[h]
	\centering
	\includegraphics[width = \linewidth]{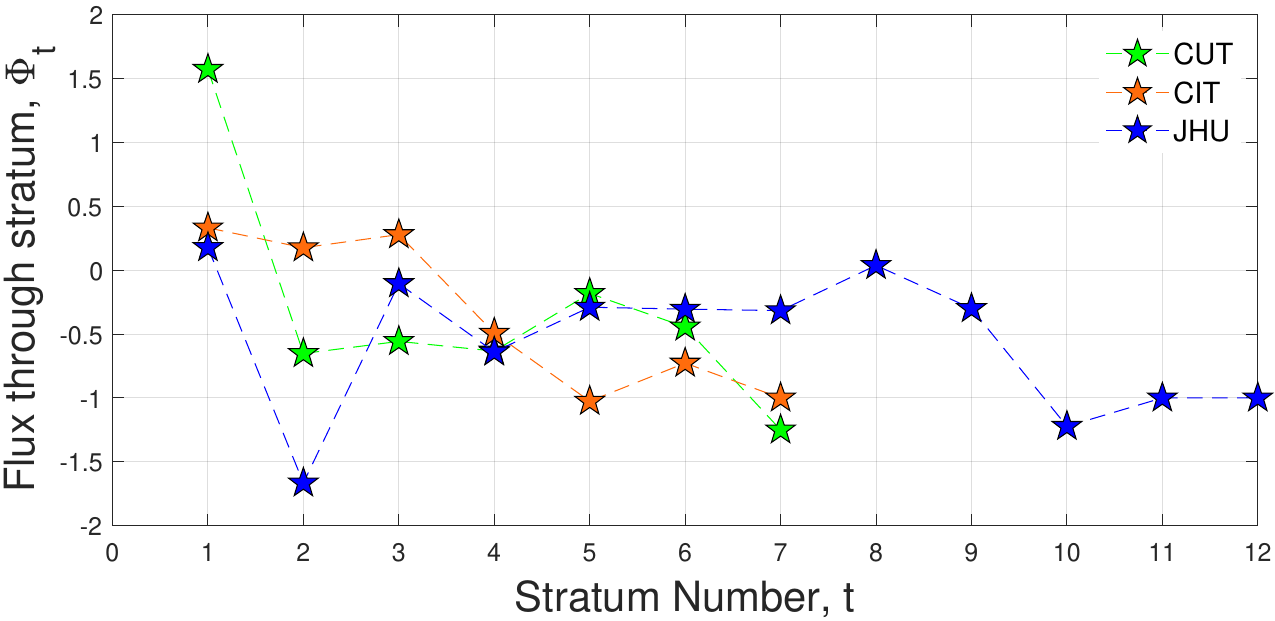}
	\caption{The flux through stratum versus the stratum number for CUT, CIT, and JHU.}
	\label{fig:8}
\end{figure}

Finally, Fig.~\ref{fig:8} shows the evolution of the flux through stratum $\Phi_t$ as the stratum number increases.  The flux $\Phi_t$ tends to decrease as we climb higher strata, since introductory courses tend to emit knowledge flow by serving as prerequisites for more advanced courses, which, in turn, tend to absorb the flow. 

\section{Summary and Discussion}
\label{sec:Conclusions}

In this paper, we propose three new global CPN measures that can be used for macro-scale comparison of course-prerequisite networks and for assessing the accuracy of random graph CPN models. The new measures are: breadth, depth, and flux, which quantify 1) how wide, 2) how deep the span of knowledge provided by the academic curriculum is, and 3) the average amount of knowledge flow through the CPN.  

All three measures are invariant with respect to the transitive reduction and are based on the concept of topological stratification, a meso-scale structure defined for directed acyclic graphs that generalizes the concept of topological ordering. To illustrate the proposed measures, we compute them for the Cyprus University of Technology, the California Institute of Technology, and Johns Hopkins University~\cite{GitHub}. 

In addition to real CPNs, we also consider synthetic CPNs generated by two random graph models: the ER and KN models, which are roughly analogous to the Erd\H{o}s–R\'{e}nyi model and the standard configuration model for undirected graphs. The KN model tends to be more accurate than the ER model in approximating the breadth, depth, and flux of real CPNs. This is not surprising, since it uses more information about the real CPN it aims to model. While the ER model uses just the number of nodes and links, the KN model uses the topological stratification and in- and out-degree sequences. Nevertheless, both models are rather inaccurate and cannot serve as realistic CPN models. 

Over the years, the development and analysis of random graph models have helped to better understand the structure and function of many different real networks. Since the ER and KN models cannot serve this purpose for CPNs, this leads to the following natural question: are there any relatively simple, amenable-to-analysis, random graph models with a few parameters that can generate synthetic DAGs similar to real CPNs? The similarity between networks can be measured with respect to classical network measures (average degree, clustering coefficient, power-law exponent, etc.) as well as the new measures proposed in this paper. 

Course-prerequisite networks originate and evolve in a highly decentralized manner: there is no single authority that decides what courses to add to the network and what links to establish between courses. These decisions are made collectively by course instructors. What local mechanisms are responsible for the formation of the global CPN structure? The answer to this question --- and the appropriate modeling approach --- may depend on how we interpret the links between nodes. Do they represent merely formal prerequisites, faculty beliefs, or objective dependencies between different areas of scientific knowledge represented by courses? We leave these important questions for future research.

\begin{acknowledgments}
We thank Gloria Brewster and Kimberley Mawhinney for providing us with a database of Caltech courses, Sergey Kushnarev and Fragkiskos Papadopoulos for their crucial assistance in obtaining course data from Johns Hopkins University and the Cyprus University of Technology, and the participants of the CMS Faculty Seminar at Caltech for their valuable feedback and stimulating discussions. 
This work was supported by the Carver Mead Discovery Grant and the Information Science and Technology (IST) initiative at Caltech.
 
\end{acknowledgments}


\begin{thebibliography}{99}
	
\bibitem{Newman2018}
Newman, M.E.J., \textit{Networks: An Introduction}, Oxford University Press, Oxford (2018).

\bibitem{NBW2006}	
Newman, M.E.J., Barab\'{a}si, A.-L. \& Watts, D.J., \textit{The Structure and Dynamics of
	Networks}, Princeton University Press, Princeton (2006).

\bibitem{Dorogovtsev2010}
Dorogovtsev, S.N., \textit{Lectures on Complex Networks}, Oxford University Press, Oxford (2010).

\bibitem{Easley2010}
Easley, D. \& Kleinberg, J., \textit{Networks, Crowds, and Markets: Reasoning about a Highly Connected World}, Cambridge University Press, Cambridge (2010).

\bibitem{Faloutsos1999}
Faloutsos, M., Faloutsos, p., Faloutsos. C., On power-law relationships of the Internet topology, \textit{ACM SIGCOMM Computer Communication Review} \textbf{29}, 251-262 (1999).

\bibitem{Pastor-Satorras2004}
Pastor-Satorras, R., Vespignani, A., \textit{Evolution and Structure of the Internet}, Cambridge University Press, Cambridge (2004).

\bibitem{Arianos2009}
Arianos, S., Bompard, E., Carbone, A., Xue, F., Power grid vulnerability: A complex network approach, \textit{Chaos} \textbf{19}, 013119 (2009).

\bibitem{Pagani2013}
Pagani, G.A., Aiello, M., The power grid as a complex network: A survey, \textit{Physica A} \textbf{392}, 2688-2700 (2013).

\bibitem{Broder2000}
Broder, A., Kumar, R., Maghoul, F., Raghavan, P., Rajagopalan, S., Stata, R., Tomkins, A., Wiener, J.,
Graph structure in the web, \textit{Computer Networks} \textbf{33}, 309-320 (2000).

\bibitem{Radicchi2012}
Radicchi, F., Fortunato, S., Vespignani, A., Citation networks, in \textit{Models of Science Dynamics: Encounters Between Complexity Theory and Information Sciences}, 233-257 (2012).

\bibitem{Martinez1991}
Martinez, N.D., Artifacts or attributes? Effects of resolution on the Little Rock Lake food web, \textit{Ecological Monographs} \textbf{61}(4), 367-392 (1991).

\bibitem{Jeong2001}
Jeong, H., Mason, S.P., Barab\'{a}si, A.-L., Oltvai, Z.N., Lethality and centrality in protein networks. \textit{Nature} \textbf{411}, 41-42 (2001). 

\bibitem{Zachary1977}
Zachary, W.W., An information flow model for conflict and fission in small groups, \textit{Journal of Anthropological Research} \textbf{33}, 452-473 (1977).

\bibitem{Borgatti2009}
Borgatti, S.P., Mehra, A., Brass, D.J., Labianca, G., Network analysis in the social sciences, \textit{Science} \textbf{323}(5916), 892-895 (2009).

\bibitem{Lusseau2003}
Lusseau, D., The emergent properties of a dolphin social network, \textit{Proceedings of the Royal Society B} \textbf{270}, S186-S188 (2003).

\bibitem{Krioukov2012}
Krioukov, D., Kitsak, M., Sinkovits, R.S., Rideout, D., Meyer, D., Bogu\~{n}\'{a}, M., Network cosmology, \textit{Scientific Reports} \textbf{2}, 793 (2012). 

\bibitem{Cunningham2017}
Cunningham, William., Zuev, K.M., Krioukov, D., Navigability of random geometric graphs in the universe and other spacetimes, \textit{Scientific Reports} \textbf{7}, 8699 (2017).

\bibitem{Krioukov2014}
Krioukov, D., Brain theory, \textit{Frontiers in Computational Neuroscience} \textbf{8}, (2014).

\bibitem{Stavrinides2023}
Stavrinides, P., Zuev, K.M.,  Course-prerequisite networks for analyzing and understanding academic curricula, \textit{Applied Network Science} \textbf{8}, 19 (2023).
	
\bibitem{Slim2014a}
Slim A., Kozlick J., Heileman G.L., Wigdahl J. \& Abdallah C.T., Network analysis of university courses, \textit{Proceedings of the 23rd International Conference on World Wide Web}, 713–718 (2014).

\bibitem{Aldrich2015}
Aldrich, P.R., The curriculum prerequisite network: modeling the curriculum as a complex system, \textit{Biochemistry and Molecular Biology Education} \textbf{43}(3), 168-180 (2015).

\bibitem{Molontay2020}
Molontay, R.,  Horv\'{a}th, N.,  Bergmann, J.,  Szekr\'{e}nyes D. \&  Szab\'{a}, M., Characterizing curriculum prerequisite networks by a student flow approach, \textit{IEEE Transactions on Learning Technologies} \textbf{13}(3), 491-501 (2020).

\bibitem{Blas2021}
Simon de Blas, C., Gomez Gonzalez, D., Criado Herrero, R., Network analysis: an indispensable tool for curricula design. A real case-study of the degree on mathematics at the URJC in Spain, \textit{PLoS ONE} \textbf{16}(3), e0248208 (2021).

\bibitem{Yang2024}
Yang, B., Gharebhaygloo, M., Rondi, H.R., Hortis, E., Lostalo, E.Z., Huang, X., Ercal, G., Comparative analysis of course prerequisite networks for five Midwestern public institutions, \textit{Applied Network Science} \textbf{9}, 25 (2024). 

\bibitem{GitHub}
GitHub: A Repository with the CPN Data,\\ 
\url{https://github.com/pstavrin/Course-Prerequisite-Networks}.

\bibitem{Aho1972}
Aho, A., Garey, M.R., Ullman, J.D., The transitive reduction of a directed graph,  \textit{SIAM Journal on Computing} \textbf{1}(2), 131-137 (1972).


\bibitem{Dunne2002}
Dunne, J.A., Williams, R.J., Martinez, N.D., Food-web structure and network theory: The role of connectance and size. \textit{PNAS} \textbf{99}(20), 12917-12922 (2002).

\bibitem{Sedgewick}
Sedgewick, R., Wayne, K., \textit{Algorithms (Fourth Edition)}. Addison-Wesley (2011).

\bibitem{Karrer2009}
Karrer, B.,  Newman, M.E.J., Random graph models for directed acyclic networks, \textit{Physical Review E} \textbf{80}, 046110 (2009).

\bibitem{GirvanNewman2002}
Girvan, M., Newman, M.E.J., Community structure in social and biological networks. \textit{Proceedings of the National Academy of Sciences} \textbf{99}(12), 7821-7826 (2002).

\bibitem{Porter2009}
Porter, M.A., Onnela, J.-P., Mucha, P.J., Communities in networks. \textit{Notices of the American Mathematical Society} \textbf{56}(9), 1082-1097 (2009).

\bibitem{Fortunato2010}
Fortunato, S., Community detection in graphs. \textit{Physics Reports} \textbf{486}(3), 75-174 (2010).

\bibitem{Holme2005}
Holme, P., Core-periphery organization of complex networks. \textit{Physical Review E} \textbf{72}(4), 046111 (2005).

\bibitem{Csermely2013}
Csermely, P., London, A., Wu, L.-Y., Uzzi, B., Structure and dynamics of core/periphery networks. \textit{Journal of Complex Networks} \textbf{1}(2), 93-123 (2013).

\bibitem{Rombach2014}
Rombach, M.P., Porter, M.A., Fowler, J.H., Mucha, P.J., Core-periphery structure in networks. \textit{SIAM Journal on Applied Mathematics} \textbf{74}(1), 167-190 (2014).

\bibitem{Alvarez2005}
Alvarez-Hamelin, J., Dall'Asta, L., Barrat, A., Vespignani, A., Large scale networks fingerprinting and visualization using the k-core decomposition. In Y. Weiss, B. Sch\"{o}lkopf \& J. Platt, eds., \textit{Advances in Neural Information Processing Systems}, vol. 18 (2005).

\bibitem{ER1959}
Erd\H{o}s, P, R\'{e}nyi, A., On random graphs. I. \textit{Publicationes Mathematicae} \textbf{6}(3-4), 290-297 (1959).

\bibitem{Bollobas1980}
Bollob\'{a}s, B., A probabilistic proof of an asymptotic formula for the number of labelled regular graphs. \textit{European Journal of Combinatorics} \textbf{1}(4), 311-316  (1980).

\bibitem{Karrer2009-2}
Karrer, B.,  Newman, M.E.J., Random acyclic networks, \textit{Physical Review Letters} \textbf{102}, 128701 (2009).
	
\end{thebibliography}
\end{document}